\documentclass[showpacs,12pt]{revtex4}%
\usepackage{amsfonts}
\usepackage{amsmath}
\usepackage{amssymb}
\usepackage{graphicx}%
\begin{document}
\preprint{Bozza 6Nov}
\title{\ Hug-like island growth of Ge on strained vicinal Si(111) surfaces}
\author{L. Persichetti, R. Menditto, A. Sgarlata, M. Fanfoni and A. Balzarotti}
\affiliation{Dipartimento di Fisica, Universit\`{a} di Roma \textquotedblleft Tor
Vergata\textquotedblright, Via della Ricerca Scientifica,1-00133 Roma, Italy.}

\begin{abstract}
We examine the structure and the evolution of Ge islands epitaxially grown on
vicinal Si(111) surfaces by scanning tunneling microscopy. Contrary to what is
observed on the singular surface, three-dimensional Ge nanoislands form
directly through the elastic relaxation of step-edge protrusions during the
unstable step-flow growth. As the substrate misorientation is increased, the
islands undergo a shape transformation which is driven by surface energy
minimization and controlled by the miscut angle. Using finite element
simulations, we show that the dynamics of islanding observed in the experiment
results from the anisotropy of the strain relaxation.

\end{abstract}
\volumeyear{year}
\volumenumber{number}
\issuenumber{number}
\eid{identifier}
\date[Date text]{date}
\received[Received text]{date}

\revised[Revised text]{date}

\accepted[Accepted text]{date}

\published[Published text]{date}

\startpage{1}
\endpage{ }

\maketitle

The formation of three-dimensional (3D) nanostructures in the
Stranski-Krastanov growth of group IV semiconductors is one of the fascinating
and complex phenomena related to heteroepitaxy
\cite{Stangl,Voigtlander,BerbezierRev,Ratto}. Among the low-index surfaces of
Si, the epitaxy of Ge on the Si(111) exhibits a relatively simple behavior
consistent with classical nucleation theory. Strain-relieving 3D islands
nucleate from fluctuations in the supersaturated wetting layer (WL) and grow
as truncated tetrahedra with \{111\} and \{113\} facets
\cite{Capellini,Voigtlander}. Within the time resolution of scanning tunneling
microscopy (STM), Ge nucleation is almost an instantaneous and homogeneous
process on the singular Si(111) surface, and only slightly correlated with
surface steps. To date, however, only a few experimental results are available
regarding highly-stepped vicinal surfaces of Si(111) \cite{BerbezierJVS,Xu}.

In this paper, we show that even a small misorientation of the substrate from
the (111) plane affects dramatically the growth dynamics of Ge relative to the
flat surface case. The highly anisotropic strain relaxation of Ge triggers the
formation of 3D structures directly from step-edge nanoprotrusions during the
unstable step-flow growth. Snapshots of the islands growth, obtained from STM
measurements, reveal an unconventional process in which the island's shape
mimics that of the elastic strain tensor as modeled by finite element (FE) simulations.

Experiments were carried out in an ultrahigh vacuum chamber (p%
$<$%
3x10$^{-11}$ torr) equipped with a STM microscope. We used Si(111) wafers
\textit{n-}doped with P ($\rho<1\Omega cm)$ with nominal azimuthal angle
$\phi$=0%
${{}^\circ}$
and polar miscut angle $\theta$ ranging between 0$%
{{}^\circ}%
$ and 9.45$%
{{}^\circ}%
$ towards the [\={1}\={1}2] direction. The uncertainty on the offcut angles
different from (9.45$%
{{}^\circ}%
\pm0.05$) $[(557)$ surface] was $\pm$0.5$%
{{}^\circ}%
$. Samples were cleaned \textit{in situ} using the following thermal pathway.
A flash heating of the substrate to 1523 K was followed by a ramp down to 1333
K during 30 s, a subsequent 2 s quench to 1103 K and a 15 min annealing at
1103 K with slow cool-down to room temperature \cite{Kirakosian}. To avoid
electromigration effects and provide a highly regular array of steps, the d.c.
heating current was parallel to the step edges along the [1\={1}0] direction
\cite{Yoshida}.%
\begin{figure}
[ptb]
\begin{center}
\includegraphics[
height=4.2661in,
width=3.3927in
]%
{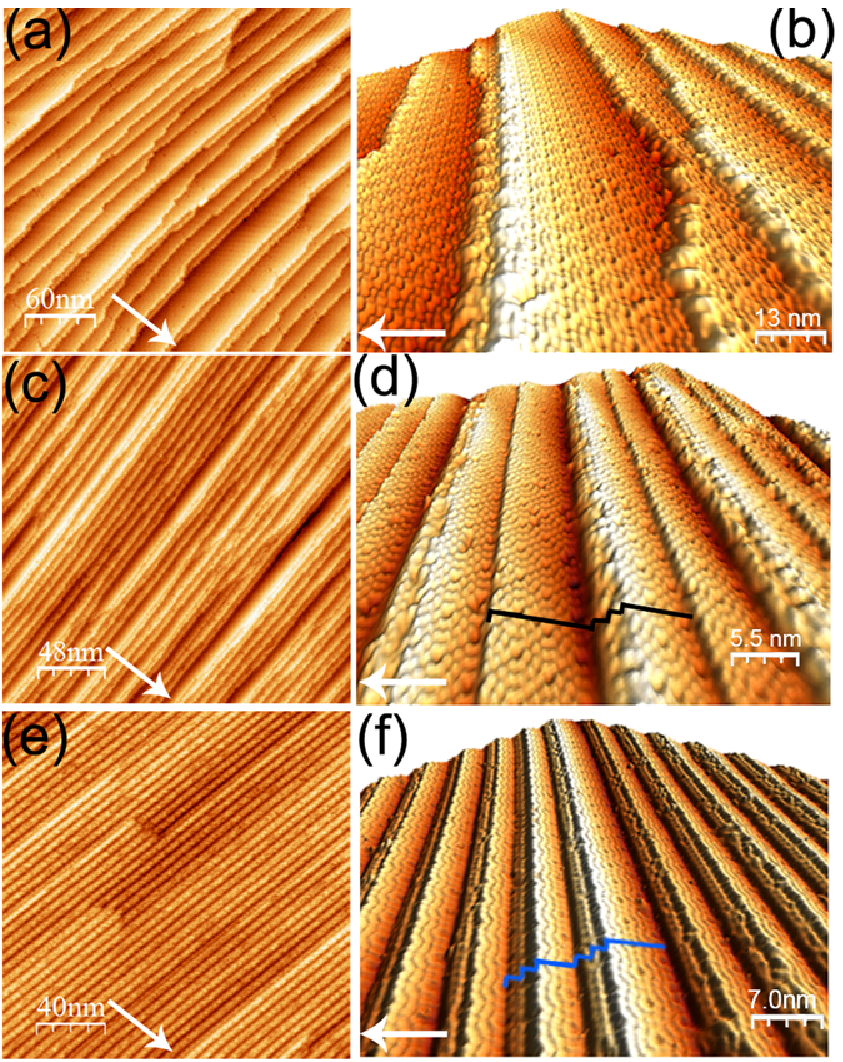}%
\end{center}
\end{figure}

Figure 1 shows the morphological structure of the clean vicinal surfaces
obtained with the treatment described above. The typical kink density
$\rho_{k}$ was in the range of 10$^{-4\text{ }}$per lattice site. The
substrates consist of (111) terraces which are (7x7) reconstructed and have
decreasing widths as the miscut becomes higher. It is also evident in the
left-hand panels of Fig. 1 that, at larger miscuts, the steps are straighter
because of the increased step-step interaction \cite{Wang} and stiffer due to
the presence of triple layers \cite{Wei,Kirakosian,Kim}. From the
high-resolution STM images displayed in Fig.1, it can be seen that triple
steps are dominant on the 9.45$%
{{}^\circ}%
$ surface [panels (e),(f)], whereas they coexist with single steps at smaller
miscuts [panels (c),(d)]. The step-step correlation function of the (557)
surface (not shown) gives an almost perfect order at long-range \cite{Lin}. On
the mesoscopic scale, the signature of the step mixture is the characteristic
behavior of step spacings as a function of the miscut angle, reported in Fig.
3(a). This is consistent with a mixed random phase of single and triple steps,
with the density of triples increasing at large miscuts (See caption of Fig. 3
for details).%
\begin{figure}
[ptb]
\begin{center}
\includegraphics[
height=3.4212in,
width=2.2935in
]%
{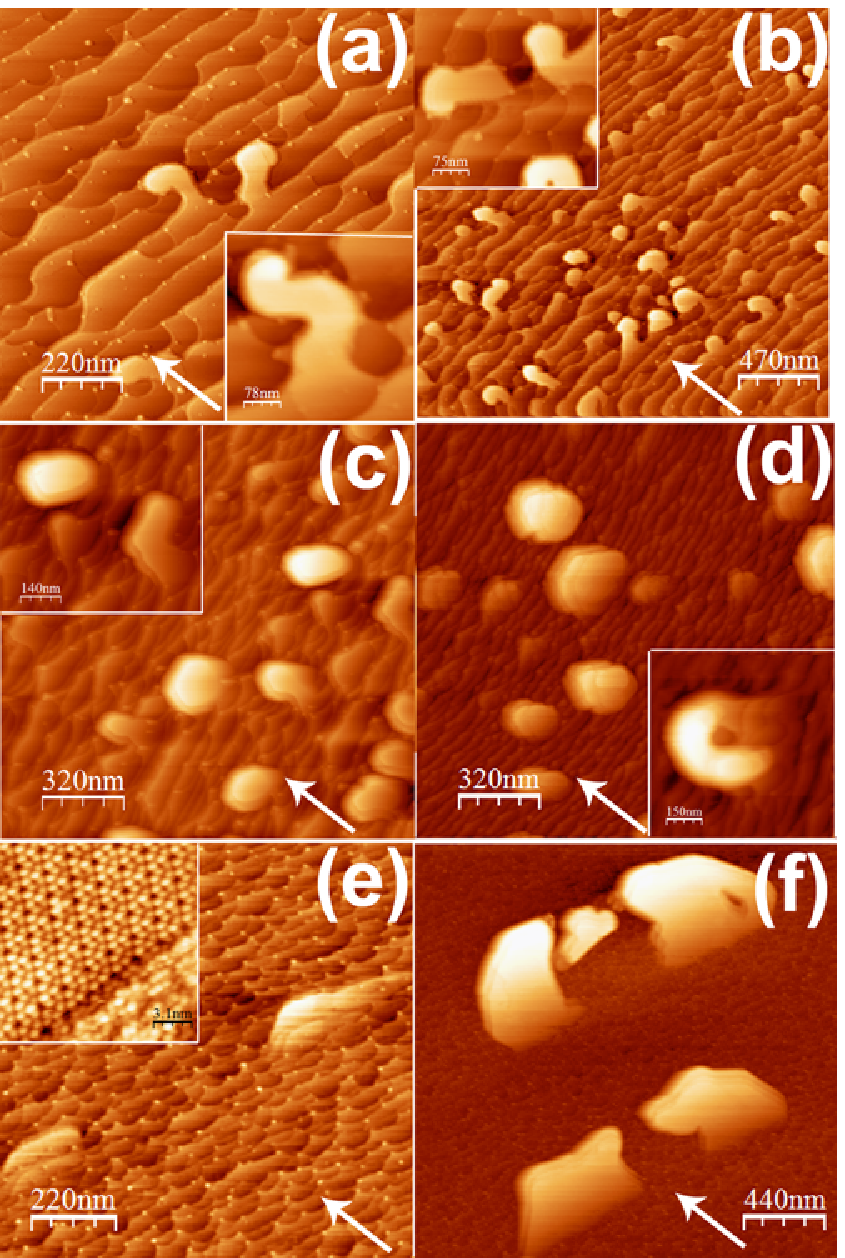}%
\end{center}
\end{figure}

On the stepped surfaces of Si(111) vicinals, Ge was deposited by physical
vapor deposition at 873 K at a constant flux of 0.1 ML/min. Figures 2(a)-(f)
show the morphological evolution induced by Ge deposition in the coverage
range 3-5 ML in which nucleation and growth of 3D islands take place on the
flat surface. During the growth, the steps initially show a characteristic
wriggling which consists of elongated protrusions, originating from extended
step-edges [Fig. 2(a,c,e)]. For larger depositions, these initially
two-dimensional nanostructures grow across the steps in the step-down
direction and become progressively taller, ultimately acquiring a 3D character
[Fig. 2(b,d,f)]. The main difference from the singular surface is that elastic
relaxation promotes the transition from a merely 2D growth of step protrusions
towards a 3D growth of Ge nanodots. The strain-driven nature of the growth is
suggested by the occurrence of the (5x5) reconstruction on the growing (111)
facets of the step-edges [inset of Fig. 2(e)], since the (5x5) reconstruction
results from a significant Ge/Si intermixing. Furthermore, the (111) facet is
the main surface orientation of the 3D islands growing from the propagation of
the protrusions. Indeed, the islands have trapezoidal shapes with a dominant
(111) facet at their top and a set of steeper lateral facets with \{113\}
orientation, as indicated by the surface orientation map (SOM) \cite{SOM} in
the inset of Fig. 3(c). Since the orientation of the (111) terraces coincides
with a dominant low-energy facet \cite{Capellini,Voigtlander,Muller}, the
process of island formation is driven by strain and surface energy
minimization. Consequently, the protrusions propagate through the steps
without disintegrating into other facets, as occurs on vicinal Si(001)
substrates \cite{Lichtemberger}. While advancing through the steps, the
protrusions grow in height following the misorientation of the substrate. As
sketched in Fig. 3(b), the smaller the terrace width, the more pronounced the
height of protrusions. Since the average terrace width has a sudden drop
between 0$%
{{}^\circ}%
$ and 1$%
{{}^\circ}%
$ [Fig. 3(a)], step protrusions spread across many steps and, hence, become
effectively 3D. In contrast, on the singular surface, they are confined to the
terrace and, thus, remain two-dimensional. As a result, the formation of 3D Ge
islands on the flat (111) surface is not coupled with step meandering but
occurs via nucleation and growth on terraces among the steps [Fig. 3(d)].
Moreover, on the vicinal substrates, the 3D islands' shape is influenced by
the distinctive growth mode. Since Ge/Si islands grow from the propagation of
the (111) terraces, their height-to-width ratio \textit{r} is set by the
average surface misorientation tan($\theta$) $\approx\theta$ [Fig. 3(b)]
\cite{note1}. Therefore,\ Ge/Si islands undergo a shape transformation which
is ruled by the preferential (111) faceting and controlled by the miscut
angle.%
\begin{figure}
[ptb]
\begin{center}
\includegraphics[
height=3.4212in,
width=1.9277in
]%
{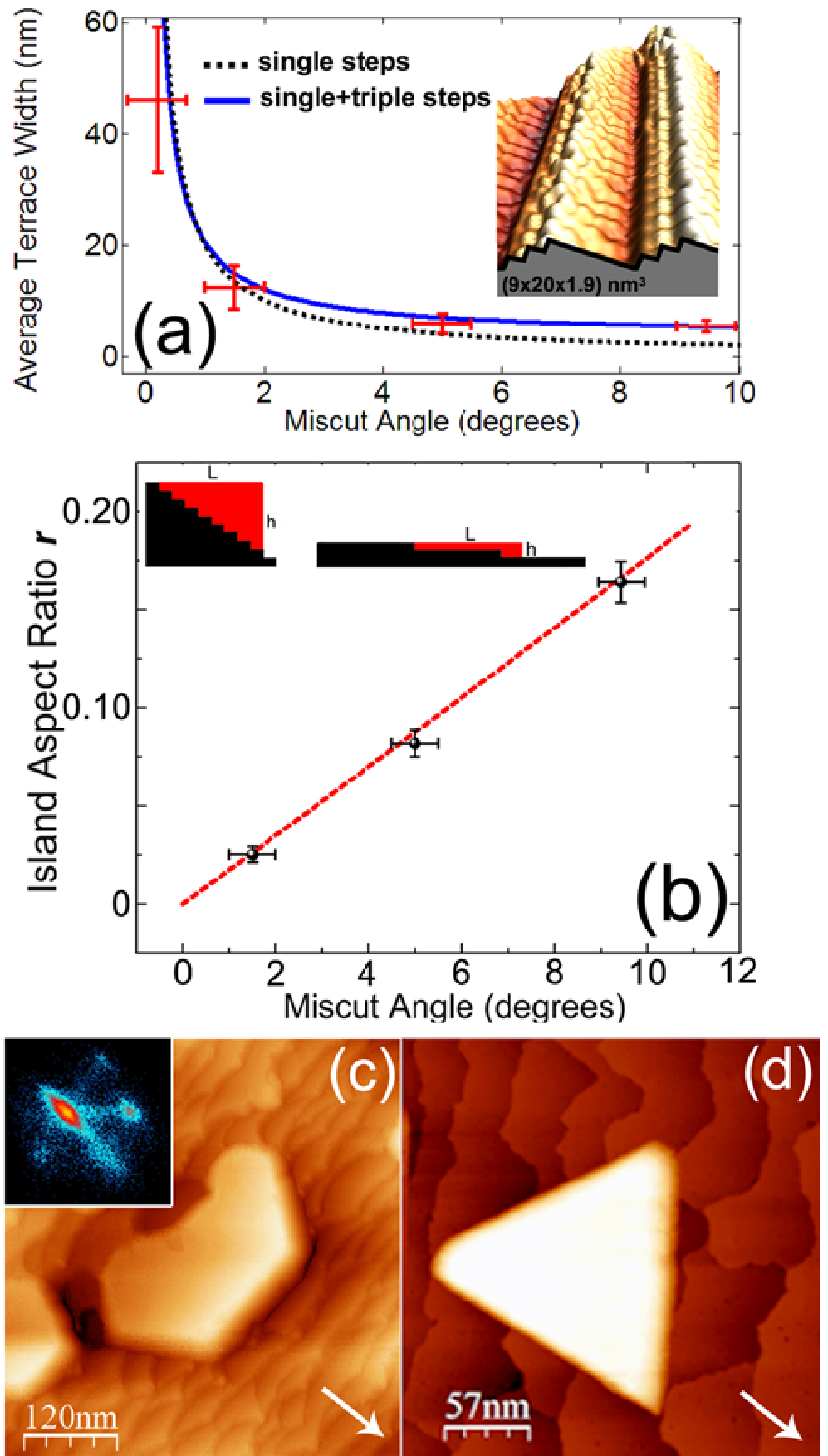}%
\end{center}
\end{figure}

Interestingly, STM images recorded at intermediate stages of growth show that
the growth mechanism of 3D islands follows a very peculiar pathway quite
different from that occurring on vicinal Si(001) surfaces \cite{Persichetti1}.
The morphology of these islands indicates a highly anisotropic growth which is
faster along the rims of the islands [Fig. 4(a-h)]. This growth mode can be
understood as a result of the anisotropic elastic relaxation of vicinal
surfaces. To this end, the equilibrium distribution of the elastic strain has
been obtained, within the continuum elasticity theory, from FE calculations
applied to the geometry of the grown islands. The equilibrium strain field
within both the island and the substrate is determined by solving the 3D
constitutive equations of elasticity for an elastic body under misfit strain
\cite{Persichetti1}. The results of such simulations are displayed in Fig.
4(i). Due to the misoriented substrate, the in-plane-strain maps are spatially
nonuniform: the relaxation of the mismatch strain ($\varepsilon_{0}$=-4\%) is
higher at the periphery of the islands than in the interior part. Due to the
anisotropy of elastic relaxation, growth is promoted along the rims by the
effective strain relief, whereas it is hindered in the highly-strained region
in the centre of the islands. Therefore, the inward growth of the islands,
outlined on the right-hand side of panel (i), effectively minimizes the strain
energy. Judging from the agreement between experiment and simulation, strain
minimization is likely to be the main driving force for the dynamic of
islanding observed on vicinal Si(111) surfaces.%
\begin{figure}
[ptb]
\begin{center}
\includegraphics[
height=3.4203in,
width=2.6005in
]%
{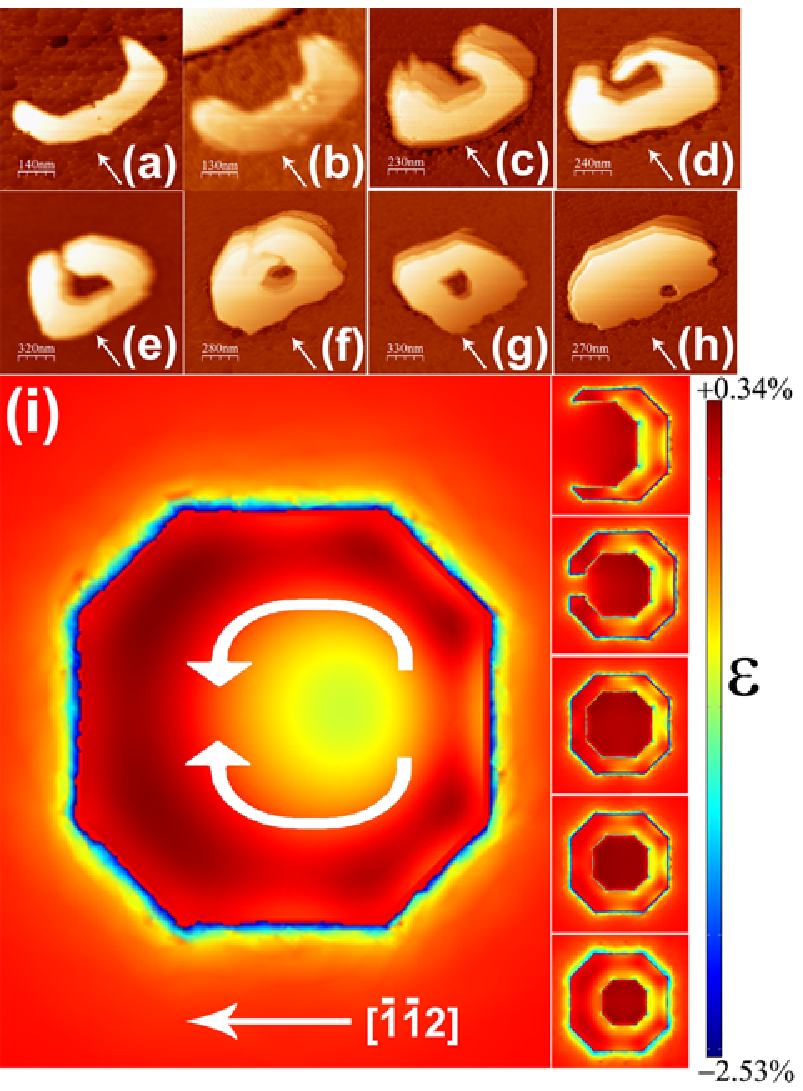}%
\end{center}
\end{figure}

In summary, we have demonstrated that that three-dimensional islanding on
vicinal Si(111) substrates occurs directly through the elastic relaxation of
step-edge protrusions during unstable step-flow growth of Ge. By simulating
the growth process with continuum elastic theory implemented in a finite
element framework, we have shown that the unconventional shape evolution of Ge
dots is a consequence of the peculiar strain field which takes place in
vicinal surfaces. This shape transformation is driven by strain energy
minimization and controlled by the miscut angle. This study contributes to a
better understanding of the role of elastic strain field in heteroepitaxy and
offers insights into the potential role of substrate vicinality for
controlling the growth of strained epitaxial nanostructures.

This work was supported in part by the Queensland Government (Australia)
through the NIRAP project "Solar Powered Nanosensors".%
\newline

{\LARGE Figure Captions}

\bigskip

Fig. 1: (color online). STM images of clean vicinal Si(111) surfaces. (a,b)
1.5$%
{{}^\circ}%
$-miscut surface. (c,d) 5$%
{{}^\circ}%
$-miscut surface. In panel (d) the coexistence of single- and triple-height
steps is highlighted. (e,f) 9.5$%
{{}^\circ}%
$-miscut surface. In panel (f) triple steps are evidenced. The arrows indicate
the [\={1}\={1}2] direction.

\bigskip

Fig. 2: (color online). STM images: (a,b) 1.5$%
{{}^\circ}%
$-miscut surface after deposition of (a) 3.6 ML and (b) 4.0 ML of Ge. (c,d) 5$%
{{}^\circ}%
$-miscut surface after deposition of (c) 3.9 ML and (d) 5.0 ML of Ge. (e,f)
9.5$%
{{}^\circ}%
$-miscut surface after deposition of (e) 3.8 ML and (f) 5.0 ML of Ge. The
arrows indicate the [\={1}\={1}2] direction.

\bigskip

Fig. 3: (color online). (a) Measured average-terrace widths of vicinal Si(111)
surfaces. The continuous line represents the expected terrace-width dependence
for a mixture of single- and triple-height steps given by $\left[  \left(
n_{s}\left(  \theta\right)  +3n_{t}\left(  \theta\right)  \right)  \left(
\tan\theta\right)  ^{-1}h\right]  ,$ where \textit{h }= 0.31 nm is the height
of a single step and $n_{s}$, $n_{t}$ are the density of single- and
triple-height steps, taken from Ref. \cite{Wei}. \ Triple steps (shown by the
STM image in the inset) increase the average step-separation compared to a
pure single-height phase (dashed curve). (b) Island aspect ratio as a function
of miscut angle. The dashed line is the average surface misorientation
tan($\theta$). (c) STM image of a Ge islands on the 5$%
{{}^\circ}%
$-miscut Si(111) surface after deposition of 4.5 ML of Ge. In the inset, the
corresponding SOM is displayed. (d) STM image of a Ge islands on the singular
Si(111) surface after deposition of 4.0 ML of Ge. The arrows indicate the
[\={1}\={1}2] direction.

\bigskip

Fig. 4: (color online). (a-h) STM images of different stages of Ge island
formation on the 1.5$%
{{}^\circ}%
$-miscut Si(111) surface. The [\={1}\={1}2] direction is indicated by arrows.
(i) FE simulations of the in-plane strain tensor $\varepsilon$ for 3D models
of Ge islands based on the experimental geometry extracted from STM images.
The white arrows indicate the direction of the island growth observed in the experiment.

\end{document}